\documentstyle[aps,prl,amsmath,amssymb]{revtex}
\draft
\begin{document}
\title{Unconventional Superconductivity
in UPd$_{\bf 2}$Al$_{\bf 3}$
from Realistic Selfconsistent Calculations}
\author{P. M. Oppeneer$^1$ and G. Varelogiannis$^{2,3}$}
\address{
$^1$Institute of Solid State and Materials Research, P.O. Box 270016,
D-01171 Dresden, Germany\\
$^2$Institute of Electronic Structure and Laser, FORTH,
P.O. Box 1527, 71110 Heraklion, Greece\\                         
$^3$Max-Planck-Institute for the Physics of Complex Systems,
N\"othnitzer Str. 38,                                              
D-01187 Dresden, Germany}
\date{\today}
\twocolumn[\hsize\textwidth\columnwidth\hsize\csname@twocolumnfalse\endcsname
\maketitle

\begin{abstract}
Realistic selfconsistent calculations of 
unconventional superconductivity in a  
heavy-fermion material are reported.      
Our calculations for UPd$_2$Al$_3$ start from 
accurate energy band dispersions that are
computed within the local spin-density functional theory
and provide Fermi surfaces
in agreement with experiment.
Using physically motivated, realistic
pairing potentials it is shown that
the superconducting gap has two lines of nodes
around the $z$-axis, thus exhibiting $d$-wave symmetry in 
the $A_{1g}$ representation of the 
$D_{6h}$ point group. 
Our results suggest that in a superconductor with
gap nodes, the prevailing gap symmetry is dictated by 
the constraint that {\it nodes must be as far as possible
from high-density areas}.
\end{abstract}
]


Superconductivity (SC)
in heavy-fermion (HF) materials exhibits a
fascinating complexity of phenomena 
\cite{Steglich}.
The nature of the SC state is mostly anticipated to be
unconventional, i.e., an additional symmetry is broken in the SC state,
and the order parameter is therefore not the conventional spin singlet
$s$-wave type 
\cite{SigristUeda,Annett}.
The identification of the order parameter symmetry is 
an essential
issue in current studies of unconventional SC.
A common approach to address the symmetry of the order parameter is through
investigations of the SC gap, which is expected to have the
same symmetry as the order parameter.
While progress has been made for
high-$T_c$ superconductors (HTSC),
the identification of the gap symmetry remains
an open question for most HF materials.
Unfortunately, the eventual identification of the gap
symmetry in a HF compound
is not sufficient to identify the pairing mechanism.
Additional elements are necessary and
in the case of HF materials, it was suggested that spin-fluctuations~-
either in the itinerant 
\cite{mathur98,saxena00}
or localized limit as magnetic excitons 
\cite{coleman01,SatoNat}~-
may replace
the phonons as mediators of the pairing.

In order to identify the gap symmetry and clarify 
its relationship to 
the pairing mechanism,
selfconsistent solutions of the SC gap equations with
physical momentum dependent pairing potentials
and accurate band structures are unavoidable.
Such type of calculations have been achieved only
recently for HTSC and ruthenates where
usually two-dimensionality is assumed and 
a tight-binding dispersion of 
one or two electron
bands is considered 
\cite{smallq1,temmerman96,Monthoux,litak01}.
In the case of HF's the situation is much more
complex. In addition to the three-dimensionality,
we must deal with several anisotropic bands
that cross the Fermi level ($E_F$) producing numerous
highly anisotropic Fermi surface (FS) sheets.
Due to the complicated structure of the bands
in HF materials, simple
tight-binding fits are impossible to obtain.
As a consequence, no
realistic selfconsistent calculations
of SC have been achieved so far in HF
materials.

Here we report 
for the first time
a direct combination of relativistic
band-structure
calculations with realistic
selfconsistent calculations of SC                 
in a HF material.
We focus on UPd$_2$Al$_3$ which is a
fascinating HF superconductor having
a moderately large specific heat
coefficient $\gamma = 140$\,mJ/mol\,K$^2$, and a
relatively high critical temperature
$T_c = 2$\,K   
\cite{geibel91}.
It orders antiferromagnetically
below $T_N = 14.3$\,K with an ordered moment of 0.85\,$\mu_B$/U-atom
\cite{krimmel92},
which is large compared to the moments of other HF superconductors
as, e.g., UPt$_3$, which has a moment of only 0.03\,$\mu_B$.
A further anomalous feature is that
the antiferromagnetic (AFM) order coexists with SC below 2\,K.
Coexistence of magnetism and SC was observed for other
materials, e.g., containing $4f$ elements,
but in those cases the magnetism is due to the localized $4f$ electrons
far from $E_F$,             
whereas the SC is carried by itinerant electrons at $E_F$.           
Conversely, for UPd$_2$Al$_3$ most of the recent 
studies reveal that
the SC, magnetic order, as well as HF behavior {\it all} involve
the uranium $5f$ states \cite{knopfle96,Bernhoeft2}.
Moreover, some striking similarities with cuprate
HTSC emerged recently.
Neutron scattering experiments 
\cite{metoki98,bernhoeft98}
have reported a distinct resonant mode that develops below $T_c$ as in
HTSC. 
Tunneling spectroscopy experiments 
\cite{jourdan99} 
in the SC state
reported peculiar dip-hump features in the tunneling spectrum,
similar to those observed in HTSC.

We solve selfconsistently the simplified 
gap equation 
starting from energy
dispersions which we computed within the framework of density-functional
theory in the local spin-density approximation (LSDA).
In our computational approach
we adopt several assumptions.
First, we approximate the multiband gap equation \cite{suhl59} 
by an effective singleband equation. 
This is justified because the bands at $E_F$ have mainly
$5f$ character\cite{knopfle96}, and there is thus no need to discriminate 
different band characters. 
The pairing potential is therefore independent of the band character.
In the multiband Hamiltonian \cite{suhl59} 
the fermion operators for the different
band characters  become thereby identical, and an effective singleband
Hamiltonian, with the energy dispersion 
$\epsilon_{\bf k}$ replaced by $\sum_n \epsilon_{n\bf k}$ ($n$ is the
band index), results.
We obtain consequently one gap symmetry defined for 
the whole FS. This corresponds to the experimental studies
that identify a global gap symmetry and not one for every FS sheet
\cite{Bernhoeft2,jourdan99,hessert97}.
Second,  since it is experimentally
firmly established that UPd$_2$Al$_3$
is a spin singlet superconductor 
\cite{NMR},
we explore in our calculations only
singlet states.
Third, we adopt an {\it adiabatic} approach in our investigation
of the SC order parameter.
This means that we neglect the renormalization of the AFM bands
by the SC, which is justified because the magnitude of
the energy scale of SC is an order of magnitude smaller than that
of the AFM order.
Fourth, in UPd$_2$Al$_3$ there exist
certainly many-particle effects that are
esponsible for the high effective mass, which can cause a
renormalization of the energy bands. Such processes are not
taken into account in our LSDA calculations.
Previous studies demonstrated a very close correspondence
of the measured
\cite{inada94} 
and calculated 
\cite{knopfle96}
de Haas-van Alphen frequencies, and consequently
the FS of UPd$_2$Al$_3$ is accurately
described by the LSDA energy dispersions. Since the
relevant physics for SC happens in the
vicinity of the FS, the LSDA bands are the appropriate
starting point for the study of the gap symmetry.
Moreover,  we focus here on a {\it qualitative}
analysis of the gap symmetry avoiding  quantitative
aspects of SC  for which the
incorrect values of the effective masses
in our LSDA scheme will indeed have some influence.

We computed the energy bands of UPd$_2$Al$_3$ with the relativistic 
augmented-spherical-wave method 
\cite{williams79}. 
The resulting LSDA
band structure of UPd$_2$Al$_3$ along
characteristic symmetry lines is shown in Fig.~1         
for the paramagnetic and AFM state.
The FS resulting from our AFM bands
consists of four types of sheets,
and is practically identical to that computed previously \cite{knopfle96},
as is also the ordered AFM moment of 0.81~$\mu_B$, which  
agrees with the experimental moment of 0.85~$\mu_B$
\cite{krimmel92}.

The simplified gap equation in the $T=0$ regime is
\begin{equation}
\Delta_{\bf k}= \frac{1}{2}
\sum_{\bf k'} V({\bf k},{\bf k'}) \Delta_{\bf k'}
( X_{\bf k'} + \Delta_{\bf k'}^2)^{-1/2} ,
\end{equation}
where $V({\bf k},{\bf k'})$ is the pairing potential,
and $X_{\bf k'} =(\sum_n \epsilon_{n\bf k})^2$ 
is the superposition of {\it all}
the relevant bands (dotted lines in Fig.~1). 
The momentum summation 
is over the
full Brillouin zone (BZ). 
An unphysical property of Eq.~(1) may occur when summed
energies are zero, but not the individual energies. 
To avoid such situation we approximate $X_{\bf k} \approx 
\sum_n \epsilon_{n\bf k}^2$. We verified that this 
approximation did not affect the resulting gap symmetry.
We consider two types of pairing kernels
that were shown to produce unconventional
order parameters.
The small-${q}$ pairing
potential adopts  at small wavevectors
the following form in momentum space
$
V({\bf k, k'})=$ $\frac{-V} {{\bf q}_c^2+({\bf k}-{\bf k'})^2}+
$ $\mu^*({\bf k}-{\bf k'})$.
This kernel is characterized by a smooth momentum cut-off ${\bf q}_c$
which selects the small-{q} processes in the
attractive part 
\cite{smallq1,Abrikosov,smallq1a,Weger,Leggett}.
At larger wavevectors the repulsive
Coulomb pseudopotential $ \mu^*({\bf k}-{\bf k'})$ prevails.
This type of attractive interaction
may be due to phonons if screening with short range
Hubbard-like Coulomb terms is involved
\cite{smallq1a}.
It has been claimed that this type of kernel 
may account for unconventional SC        
in HTSC 
\cite{smallq1,Leggett} and
in other materials, including HF's
\cite{smallq1a}.
Agterberg {\it et al}. 
\cite{Agterberg}
have recently suggested for
HF superconductors 
a multipocket FS with an attractive
intrapocket and repulsive interpocket potential.
Such situation results naturally from a potential
dominated by small-{q} attractive pairing as the one
considered here,
if the FS contains multipockets.
We also considered the case of pairing mediated by 
{\it spin-fluctuations} 
using a phenomenological
Millis-Monien-Pines pairing potential  
\cite{millis90}
which has the form
$V({\bf k, k'})=\frac{V}{{\bf q}_c^2+({\bf k}-{\bf k'}-{\bf Q})^2}$
where ${\bf Q}=(0,0,\pi/2c)$ in UPd$_2$Al$_3$ ($c$ is the $c$-axis      
lattice constant).
This type of pairing has been suggested for
virtually all unconventional SC's, including UPd$_2$Al$_3$ 
\cite{HuthEPJ,Bernhoeft2}.

To obtain selfconsistent gap solutions
with such momentum dependent kernels
we have used a Fast-Fourier-Transform technique.
The problem is solved iteratively on a symmetric             
part of the BZ
that we discretize with a
$128\times 128 \times 128$ momentum grid. 
For each point of our 3-D grid we have
our LSDA energy bands as an input.
Within our procedure the momentum space problem is fully
resolved numerically without any simplification or          
bias on the resulting gap symmetry. 

In UPd$_2$Al$_3$, NMR measurements revealed a singlet
SC state 
\cite{NMR} 
with a gap of even parity.
Because UPd$_2$Al$_3$ and thus our LSDA bands obey the hexagonal
$D_{6h}$ point group symmetry, the only even parity
accessible gap
states with nodes should transform according to
the irreducible representations of
$D_{6h}$
\cite{Yip} 
shown schematically in Fig.~2.
As a first step in our iteration scheme,
the initial gap configuration is chosen randomly and the system
is totally free to converge within the iteration cycle towards
the most favorable representation.
In a second step we bias the system
by adopting as initial gap configuration each of these
representations and try to force
convergence towards the chosen representation.
In that way, we check the
ground state
character of the solution obtained in the first step.
The precision of our numerical calculation is such that
it allows to 
distinguish which representation corresponds to the lowest
free energy.
The lowest-in-energy representation, e.g., has a free energy  
that is 10\% less than the next higher one.
The second step 
in addition 
allows us to
identify any eventual degeneracy between the
various accessible gap symmetries.

The central result of our calculations is that
the only accessible gap states with nodes in UPd$_2$Al$_3$
{\it transform according to the fully symmetric irreducible
representation $A_{1g}$}. This is true {\it irrespective
of the precise momentum structure of the pairing kernel}.
We show in Fig.~3 some examples of our
selfconsistently obtained gap functions
in the plane $ {\rm L} -{\rm A} - {\rm L} - {\rm L} - {\rm M} 
- {\rm L}$ which contains
the $z$-axis (the $ {\rm A} - \Gamma - {\rm A}$ axis) and in the
plane obtained by a $\pi/6$ or $\pi/2$ rotation around the $z$-axis
(the plane $ {\rm H - A - H - H - K - H}$). All
selfconsistent
solutions shown in fig.~3 have
two lines of nodes practically perpendicular to the ${\rm A} - \Gamma - 
{\rm A}$ axis, thus
belonging in the $A_{1g}$ representation.
{\it No other representation possessing nodes was accessible} for all
parameters in the pairing kernels that we have investigated.
To study the influence of the coexisting AFM
order on SC we have made calculations using both paramagnetic and
AFM bands.
In both paramagnetic and AFM cases, the only solutions with
gap nodes are of $A_{1g}$ type.
Note that
our results with the spin-fluctuations kernel
confirm the model
analysis by Huth {\it et al}. 
\cite{HuthEPJ}.

There is overwhelming experimental evidence that
indeed UPd$_2$Al$_3$
has a gap with the
$A_{1g}$ node structure. 
Recent tunnel measurements along the $z$-axis showed the absence of a
node in this direction 
\cite{jourdan99}.
However,
the presence of nodes is definitely established in UPd$_2$Al$_3$, and
since $A_{1g}$ is
the only even parity gap with nodes which is nodeless
along the $z$-direction, it was deduced {\it a posteriori} that the
gap symmetry must be $A_{1g}$ type 
\cite{jourdan99}. 
In addition, the observation of
a spin-fluctuations peak below $T_c$ at $(0,0,\pi/2c)$
\cite{metoki98,bernhoeft98} 
would be forbidden by
the involved
coherence factors unless the gap changes
sign along the $z$-axis as in
the $A_{1g}$ representation 
\cite{comment}.
Finally, measurements of the angular dependence
of the critical field indicate two lines of
nodes perpendicular to the $z$-axis 
\cite{hessert97} 
again suggesting
the $A_{1g}$ representation. UPd$_2$Al$_3$
is perhaps the only HF superconductor for which
such a clear picture of the nodal structure of the gap is obtained
from the experiments. We consider the agreement of our results
with the experiments as strong support of our calculations.

The surprising robustness of the $A_{1g}$ solution
results from the following general rule:
If a system must choose a gap symmetry with nodes
because of the repulsive effective interaction
at large wavevectors (short distances)
it chooses {\it the representation that
has the minimum number of nodes as far as possible from the
high-density areas}.
In fact, the system must  {\it maximize} the condensation
free energy and this is obtained when there is a gap
in the high-density areas and the node (gapless)
areas are kept minimal.
Examining the paramagnetic bands of UPd$_2$Al$_3$ (Fig.~1b)
we see that the high-density areas near
the FS due to saddle points in the band dispersions
are found essentially near the A point and
near the H point. The $A_{1g}$ representation
prevails because it is the
only one {\it without a node near the A point}.
In the case of the AFM bands (Fig.~1c), the saddle points
of the bands near the FS are found
essentially in the $z=0$ plane (near the $\Gamma$ point,
along the $\Gamma-\rm K$, $\rm K-M$ and $\rm M-\Gamma$ 
symmetry lines)
and near the $\Gamma$ point along $\Gamma - \rm A^{\prime}$
 ($\rm A^{\prime} = A/2$).
The high-density areas in the $z=0$ plane exclude
the $B_{1g}$, $E_{1g}$ and $B_{2g}$ representations which
have a line node in the $z=0$ plane.
From the remaining $A_{1g}$, $E_{2g}$ and $A_{2g}$
representations, $A_{1g}$ is the only one without
a line of nodes parallel to the $\rm A-\Gamma-A$ axis
which would cross the high-density areas at
$\Gamma$ and near $\Gamma$ along $\rm \Gamma-A^{\prime}$.
Also, due to the AFM symmetry a high density occurs again at
$\rm A$, as $\rm A$  is equivalent to $\Gamma$.
The node perpendicular to the $\rm \Gamma-A$ path
in the $A_{1g}$ representation is 
close to $\rm A^{\prime}$ and therefore                
does not cross the high-density areas near $\Gamma$ and 
$\rm A$.
Note that $A_{2g}$, $B_{1g}$ and $B_{2g}$ are
also handicapped by the fact that they have more
nodal areas.

To stress the generality of this argument we reconsidered the
case of cuprate HTSC.
Saddle points produce
high-density areas in the vicinity of $(0,\pi/a)$ and symmetry related
areas ($a$ is the lattice constant in the basal plane). 
The $d_{x^2-y^2}$ gap symmetry corresponds to nodes along the
$(\pi/a,\pi/a)$ direction as far as possible from the
high-density $(0,\pm\pi/a)$ and $(\pm\pi/a,0)$ areas.
Using both types of pairing kernels as adopted 
for UPd$_2$Al$_3$ (now ${\bf Q}=(\pi/a,\pi/a)$)
we were never able to obtain
a $d_{xy}$ solution for which the nodes cross
the high-density
$(0,\pi/a)$ areas {\it no matter the details of the
interaction.}


In conclusion, we have computed the first selfconsistent 
solutions of SC in the HF material 
UPd$_2$Al$_3$.
Our calculations demonstrate that the only even parity
accessible gap symmetry with nodes transforms according to 
the $A_{1g}$ irreducible representation of the 
$D_{6h}$ point group, independent of whether 
the pairing potential 
is dominated by small-q attractive processes or by 
spin-fluctuations. The robustness of the $A_{1g}$ representation
is analyzed
to be due to the     
presence of high-density regions 
of the relevant bands in the vicinity
of the $\Gamma$ and $\rm A$ points, that preclude node
formation at these points.
As a general rule we obtain that nodes must be 
as far as possible from high-density areas in the phase space.

We are grateful to H. Adrian, E.\,N. Economou,
P. Fulde, M. Huth, M. Lang, P. Thalmeier, N. Sato and
F. Steglich for illuminating discussions. We thank M. Huth for the
figure of the irreducible representations of the $D_{6h}$ group.



\begin{figure}[tbp]
\caption{
({\bf a}) The hexagonal Brillouin zone
with high symmetry points. In the AFM phase the BZ
is reduced by a factor two along the $z$-axis 
(denoted by the primed letters).
({\bf b}) The energy bands along high symmetry directions
in the paramagnetic, and ({\bf c}) the AFM phase of UPd$_2$Al$_3$.
The relevant bands for superconductivity
are shown by the dotted lines. 
}
\label{fig1}
\end{figure}

\begin{figure}[tbp]
\caption{
The even parity irreducible representations
of the $D_{6h}$ point group. While in principle all are accessible
for the gap, only the $A_{1g}$ results from our selfconsistent
gap calculations.
}
\label{fig2}
\end{figure}

\begin{figure}[tbp]
\caption{
(color) Some examples of the selfconsistent gap solutions
in characteristic planes 
(with sides indicated)
that contain the 
$\Gamma$ point (in the center)
and the $z$-axis 
($\rm A-\Gamma-A$ axis). The thick black lines
show the nodes and in the blue areas the gap is negative.
All nodes cut the $\rm A - \Gamma - A$ axis as in the $A_{1g}$
representation. We show here results obtained with
small-q pairing and AFM energy bands,       
({\bf a} and {\bf b}), spin-fluctuations pairing and AFM bands ({\bf c}), 
and small-q pairing and paramagnetic bands ({\bf d}). In ({\bf d}) the
sign of the gap is inverted to visualize better the variation of the
gap.
}
\label{fig3}
\end{figure}

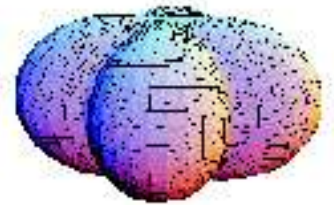
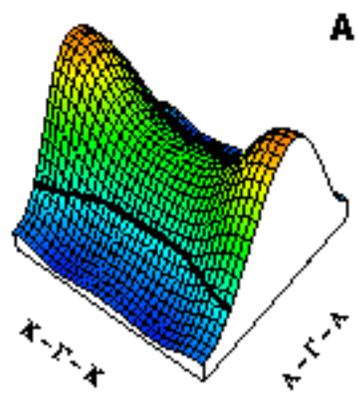
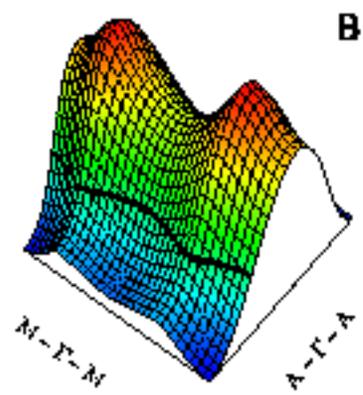
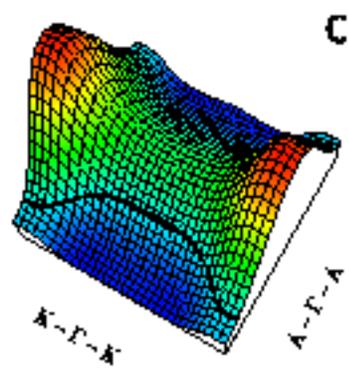
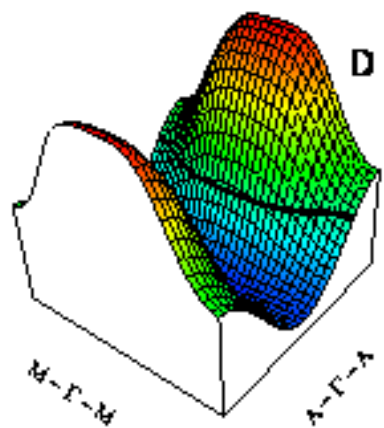
\end{document}